\documentclass[paper]{JHEP3}
\usepackage{epsfig}
\usepackage{amsmath}
\usepackage{amssymb}
\usepackage{amsbsy}
\usepackage{enumerate}
\usepackage{bbm}
\usepackage{url}

\usepackage[hang,sf]{subfigure}

\newenvironment{enumeratealpha}{\begin{enumerate}[a{\textup{)}}] }
                               {\end{enumerate}}

\def\beq{\begin{equation}}
\def\beqn{\begin{eqnarray}}
\def\eeq{\end{equation}}
\def\eeqn{\end{eqnarray}}

\newcommand\HERWIG{{\tt HERWIG}}

\newcommand\PYTHIA{{\tt PYTHIA}}

\newcommand\VBFNLO{{\tt VBFNLO}}
\newcommand\MCFM{{\tt MCFM}}

\def\lq{\left[} 
\def\rq{\right]}

\def\({\left(} 
\def\){\right)}

\newcommand\sss{\mathchoice%
{\displaystyle}%
{\scriptstyle}%
{\scriptscriptstyle}%
{\scriptscriptstyle}%
}

\newdimen\hbigcirc
\newdimen\wbigcirc

\newdimen\figwidth

\newcommand\captskip{\vskip -0.7cm}

\figwidth=\textwidth

\newcommand\as{\alpha_{\sss\rm S}}

\newcommand\pt{p_{\sss\rm T}}
\newcommand\ptmin{{\pt^{\min}}}
\newcommand\kt{k_{\sss\rm T}}

\newcommand\mur{\mu_{\sss\rm R}}
\newcommand\muf{\mu_{\sss\rm F}}

\newcommand\MCatNLO{{\tt MC@NLO}}

\newcommand\CF{C_{\sss\rm F}}

\newcommand\POWHEG{{\tt POWHEG}}
\newcommand\POWHEGBOX{{\tt POWHEG BOX}}

\newcount\minutes 
\newcount\scratch 
 
\def\timestamp{%
\scratch=\time 
\divide\scratch by 60 
\edef\hours{\the\scratch} 
\multiply\scratch by 60 
\minutes=\time 
\advance\minutes by -\scratch 
---$\,$\hours:\null 
\ifnum\minutes< 10 0\fi 
\the\minutes}

\title{NLO Higgs boson production via vector-boson fusion \\
matched with shower in {\tt\bf POWHEG}}
\author{Paolo Nason\\
  INFN, Sezione di Milano-Bicocca,
  Piazza della Scienza 3, 20126 Milan, Italy\\
  E-mail: \email{Paolo.Nason@mib.infn.it}}
\author{Carlo Oleari\\
  Universit\`a di Milano-Bicocca and INFN, Sezione di Milano-Bicocca\\
  Piazza della Scienza 3, 20126 Milan, Italy\\
  E-mail: \email{Carlo.Oleari@mib.infn.it}}
\abstract{
  We present a next-to-leading order calculation of Higgs boson production in
  vector-boson fusion processes interfaced to shower Monte Carlo programs,
  implemented according to the \POWHEG{} method. 
}

\begin{document}

\section{Introduction}
\label{sec:introduction}
Higgs boson production via vector-boson fusion~(VBF) is expected to provide a
copious source of Higgs bosons in $pp$-collisions at the Large Hadron
Collider~(LHC) at CERN. It can be visualized (see
fig.~\ref{fig:born_virt}(a)) as the inelastic scattering of two quarks
(antiquarks), mediated by $t$-channel $W$ or $Z$ exchange, with the Higgs
boson radiated off the weak bosons.  It represents (after gluon fusion)
the second most important production process for Higgs boson
studies~\cite{CMS,ATLAS}.  Once the Higgs boson has been found and its mass
determined, the measurement of its couplings to gauge bosons and fermions
will be of main interest~\cite{Zeppenfeld:2000td,Duhrssen:2004cv}.  Here VBF
will play a central role since it will be observed in the
$H\to\tau\tau$~\cite{Rainwater:1998kj,Plehn:1999xi}, $H\to
WW$~\cite{Rainwater:1999sd, Kauer:2000hi} and
$H\to\gamma\gamma$~\cite{Rainwater:1997dg} channels. This multitude of
channels allows to probe the different Higgs boson couplings.  The VBF
measurements can be performed at the LHC with statistical accuracies
of the order of 5 to
10\%~\cite{Zeppenfeld:2000td}.
In addition, in order to distinguish the VBF Higgs boson signal from
backgrounds, stringent cuts are required on the Higgs boson decay products as
well as on the two forward quark jets which are characteristic for VBF. The
efficiency of these cuts has to be evaluated on the basis of the most updated
simulation tools and experimental inputs. It is typically of the order of
25\%.

In the past few years, the next-to-leading order~(NLO) corrections to Higgs
boson production in VBF have become
available~\cite{Figy:2003nv,Ravindran:2002dc, Berger:2004pca} and implemented
in fully-flexible partonic Monte Carlo programs (see, for example, the
\VBFNLO{} package~\cite{Arnold:2008rz} and the \MCFM{} code~\cite{MCFM}). The
NLO QCD corrections have been shown to be modest~\cite{Figy:2003nv}, of order
5 to 10\% in most cases, but reaching 30\% in some distributions. In
addition, scale uncertainties range from 5\% or less for distributions to
below $\pm 2$\% for the Higgs boson total cross section after typical
VBF cuts.

In addition, the dominant NLO corrections to $Hjjj$ in VBF have been computed
in ref.~\cite{Figy:2007kv}. In refs.~\cite{Andersen:2006ag,
  Bredenstein:2008tm, Ciccolini:2007ec, Andersen:2007mp}, interference of VBF
and Higgsstrahlung, electroweak corrections and loop-induced interference
effects have been computed too, and it was shown that, in the phase-space
region relevant for VBF observation at the LHC, these effects are small, so
that they can be neglected at first approximation.

In this work we merge the NLO calculation of Higgs boson production via VBF
with a shower Monte Carlo program, according to the \POWHEG{} method.  This
method was first suggested in ref.~\cite{Nason:2004rx}, and was described in
great detail in ref.~\cite{Frixione:2007vw}.  Until now, the \POWHEG{} method
has been applied to $ZZ$ pair hadroproduction~\cite{Nason:2006hfa},
heavy-flavour production~\cite{Frixione:2007nw}, $e^+ e^-$ annihilation into
hadrons~\cite{LatundeDada:2006gx} and into top
pairs~\cite{LatundeDada:2008bv}, Drell-Yan vector boson
production~\cite{Alioli:2008gx,Hamilton:2008pd}, $W'$
production~\cite{Papaefstathiou:2009sr}, Higgs boson production via gluon
fusion~\cite{Alioli:2008tz,Hamilton:2009za}, Higgs boson production
associated with a vector boson (Higgs-strahlung)~\cite{Hamilton:2009za},
single-top production~\cite{Alioli:2009je} and $Z+1$~jet
production~\cite{POWHEG_Zjet}. Unlike the \MCatNLO{}
implementation~\cite{Frixione:2002ik}, \POWHEG{} produces events with
positive (constant) weight, and, furthermore, does not depend on the
subsequent shower Monte Carlo program. It can be easily interfaced to any
modern shower generator and, in fact, it has been interfaced to
\HERWIG{}~\cite{Corcella:2000bw,Corcella:2002jc} and
\PYTHIA{}~\cite{Sjostrand:2006za} in
refs.~\cite{Nason:2006hfa,Frixione:2007nw,Alioli:2008gx,Alioli:2008tz,
Alioli:2009je}.

This paper is organized as follows.  In section~\ref{sec:implementation} we
briefly introduce the \POWHEGBOX{} package, that will be discussed in great
detail in a forthcoming paper~\cite{POWHEGBOX}.  We describe some of its new
features (that were not present in previous \POWHEG{} implementations) needed
in order to deal with Higgs production in VBF in a more efficient way. In
section~\ref{sec:results} we show our results for several typical kinematic
variables, often discussed when dealing with VBF Higgs boson production.

Since this is the first NLO plus shower implementation of Higgs boson
production in VBF, we could not make any comparison with other similar
results.  We limit ourselves to make comparisons among the pure NLO
distributions and the \POWHEG{} results showered by \HERWIG~6.510 and
\PYTHIA{}~6.4.21.  Finally, in section~\ref{sec:conc}, we give our
conclusions.

\section{The \POWHEG{} implementation}
\label{sec:implementation}
\subsection{The \POWHEGBOX}
\label{sec:powhegbox}
We have implemented VBF Higgs boson production inside the
\POWHEGBOX~\cite{POWHEGBOX}. This is an automated package that turns a NLO
calculation into a \POWHEG{} one, and whose output are events ready to be
showered by any shower Monte Carlo program, such as \HERWIG{} or \PYTHIA.
All the details and an explanation of how it works will be given in a 
forthcoming paper~\cite{POWHEGBOX}.

In order to build a \POWHEG{} implementation using the \POWHEGBOX, one has to
provide the following:
\begin{enumeratealpha}
\item A list of all flavour structures of the Born processes.
  
\item The Born squared amplitude ${\cal B}$ and the Born phase space.
  
\item \label{item:born} The color correlated ${\cal B}_{ij}$ and spin
  correlated ${\cal B}_{\mu\nu}$ Born cross sections. These are common
  ingredients in NLO calculations regularized with a subtraction method.

\item The Born color structure in the large limit of the number of colors.

\item \label{item:virtual} The finite part of the virtual corrections ${\cal
  V}_{\rm fin}$, computed in dimensional regularization or in dimensional
reduction.
  
\item The list of all the flavour structures of the real processes.
  
\item \label{item:real} The real matrix elements squared for all relevant
  partonic processes.

\end{enumeratealpha}
The \POWHEGBOX{} then finds all the singular regions, builds the soft and
collinear counterterms and the soft and collinear remnants, and then 
generates the radiation with the \POWHEG{} Sudakov form factor.

\begin{figure}[thb] 
\centerline{ 
\subfigure[tree-level diagram]{
\epsfig{figure=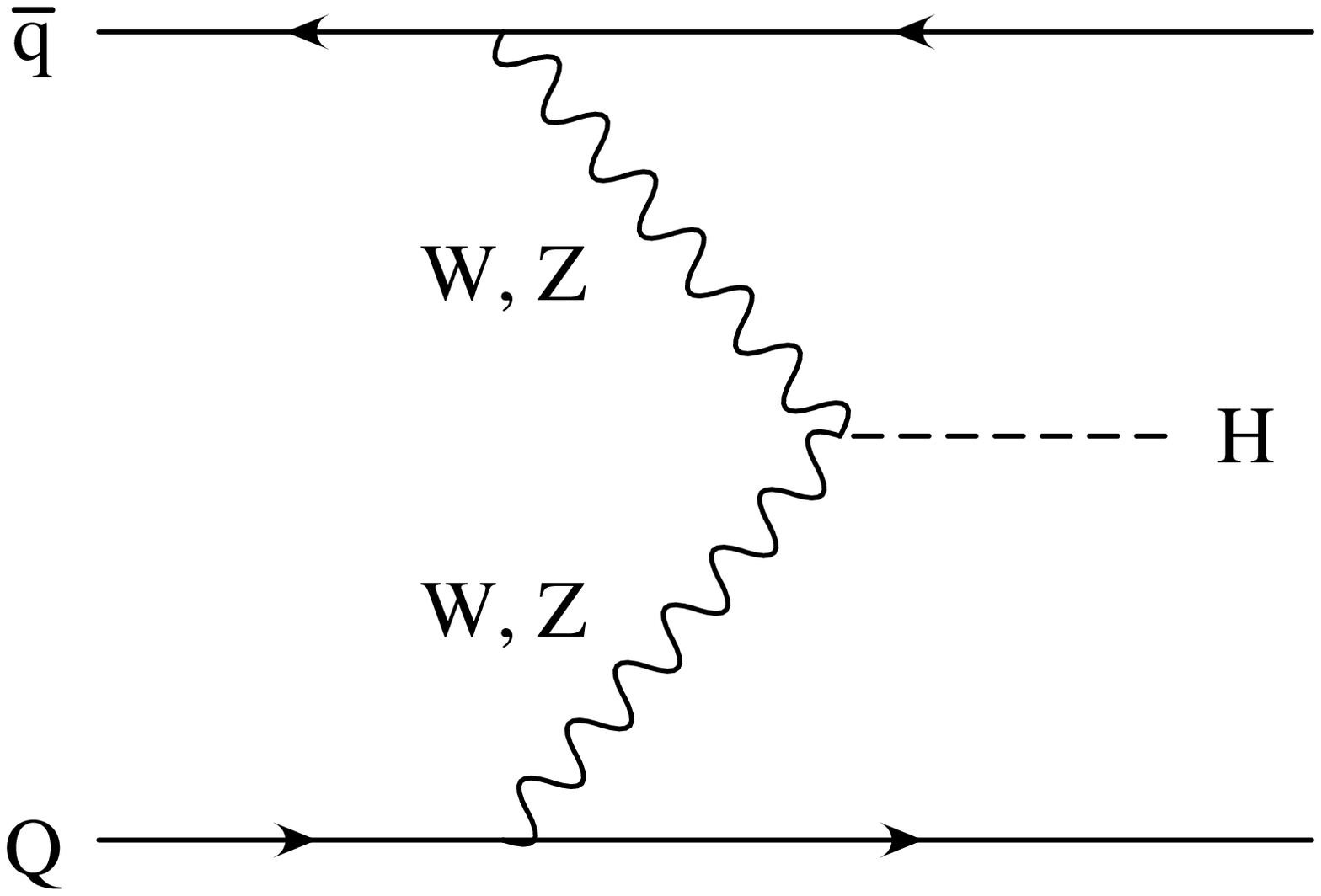,width=0.35\textwidth,clip=}}
\qquad
\subfigure[virtual diagram]{ 
\epsfig{figure=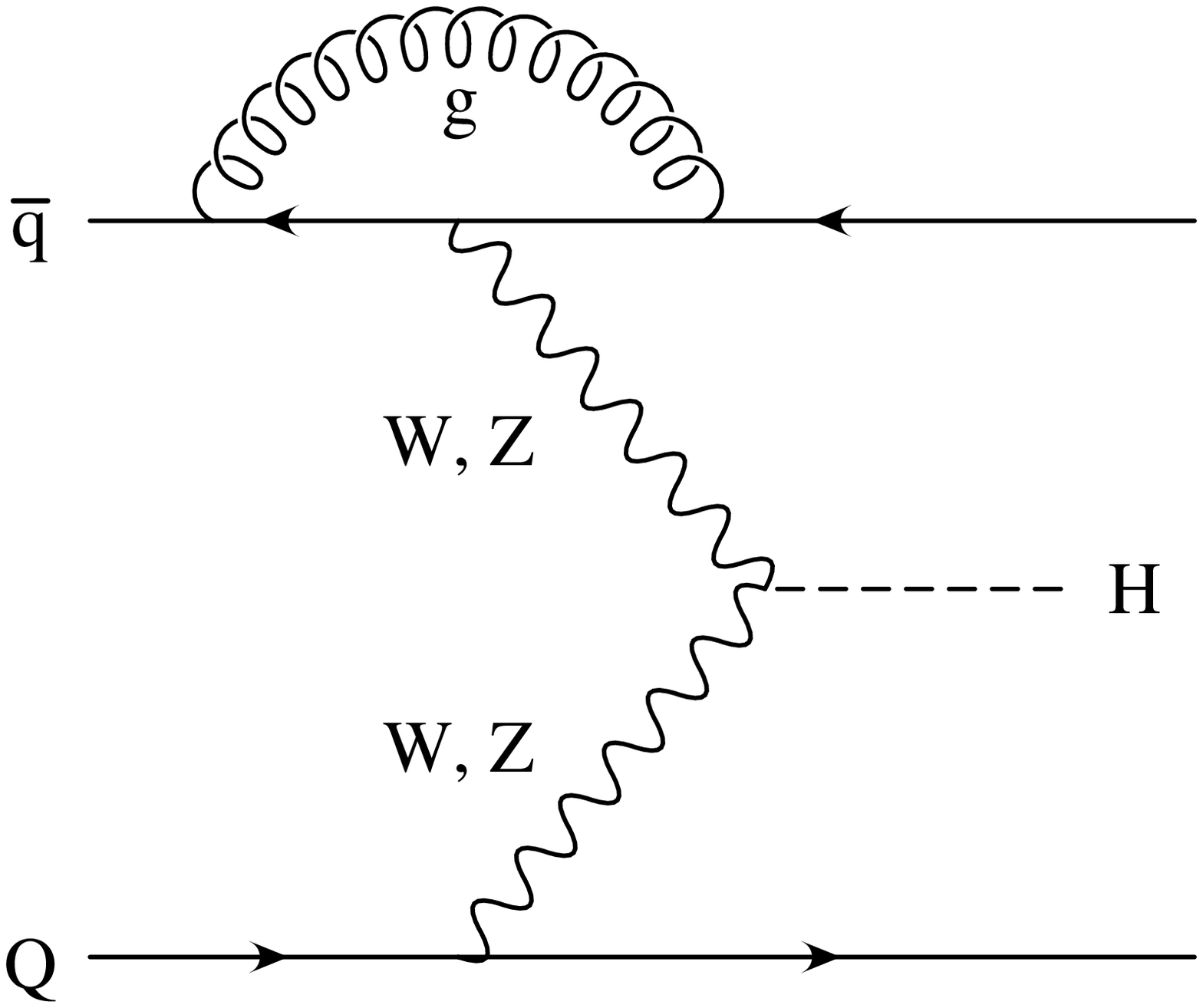,width=0.35\textwidth,clip=}}
} 
\caption{ Feynman graphs contributing to $\bar q(1)\, Q(2) \to  H(3)\,
\bar q(4)\, Q(5)$ at tree level~(a) and including virtual corrections to
the upper quark line~(b).}
\label{fig:born_virt}
\end{figure}
At lowest order, Higgs boson production via vector-boson fusion is
represented by a single Feynman graph, like the one depicted in
fig.~\ref{fig:born_virt}(a) for $\bar q(1)\, Q(2) \to  H(3)\, \bar q(4)\,
Q(5)$. The numbers in parenthesis represent a collective index that
identifies flavour, color, spin and momentum of the corresponding particle.
Strictly speaking, the single Feynman graph picture is valid
only for different quark flavours on the two fermion lines. For identical
flavours, annihilation processes, like $\bar q q\to Z^*\to ZH$ with
subsequent decay $Z\to \bar q q$, or similar $WH$ production channels,
contribute as well. In addition, for $qq\to Hqq$ or $\bar q\bar q\to H\bar
q\bar q$, the interchange of identical quarks in the initial or final state
needs to be considered in principle. However, in the phase-space regions
where VBF can be observed experimentally, with widely separated quark jets of
very large invariant mass, the interference of these additional graphs is
strongly suppressed by large momentum transfer in the weak-boson
propagators. Furthermore, color suppression makes these effects
negligible. In the following we systematically neglect any identical-fermion
effects.

\begin{figure}[thb] 
\centerline{ 
\subfigure[final-state radiation]{
\epsfig{figure=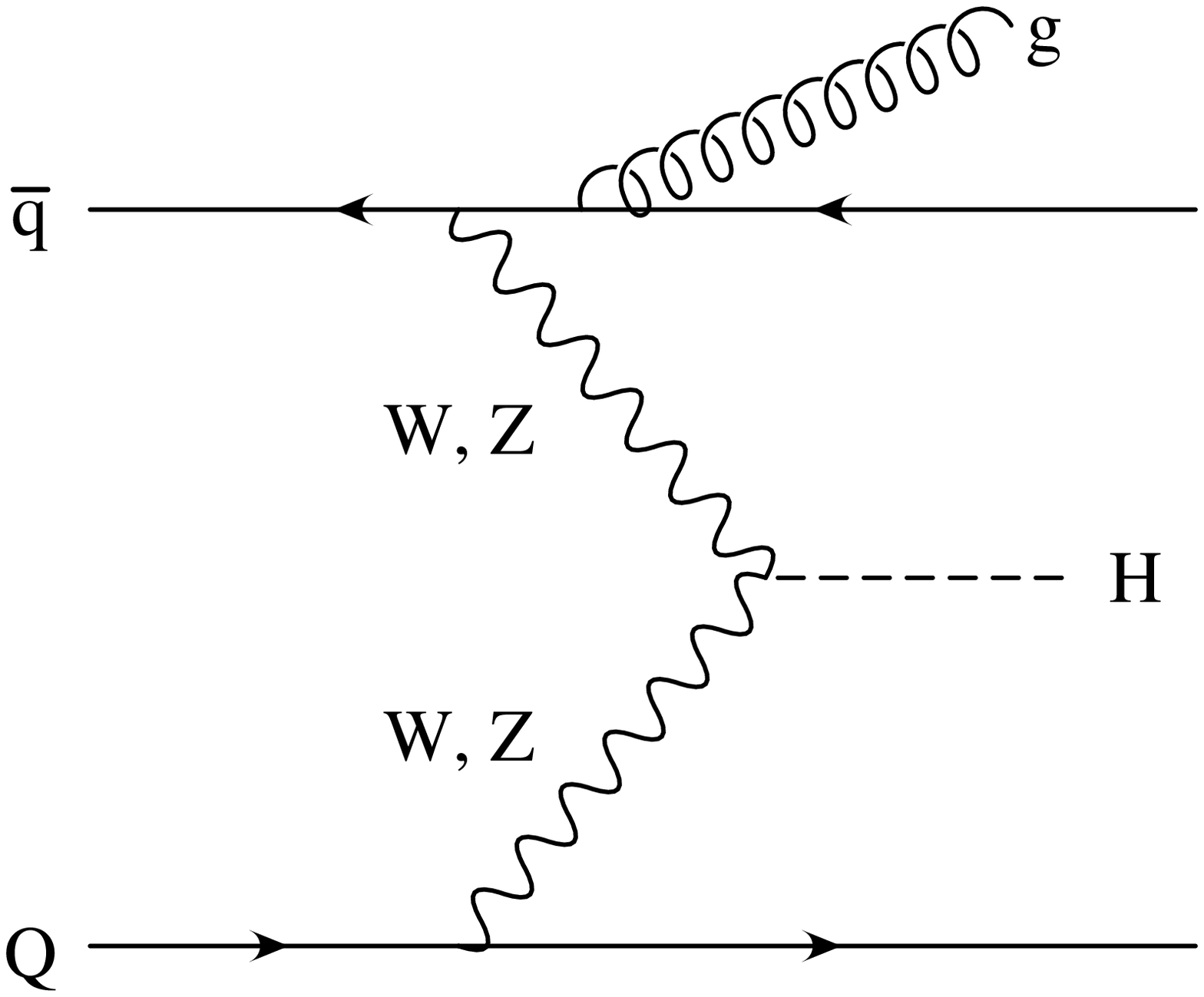,width=0.35\textwidth,clip=}}
\qquad
\subfigure[initial-state radiation]{ 
\epsfig{figure=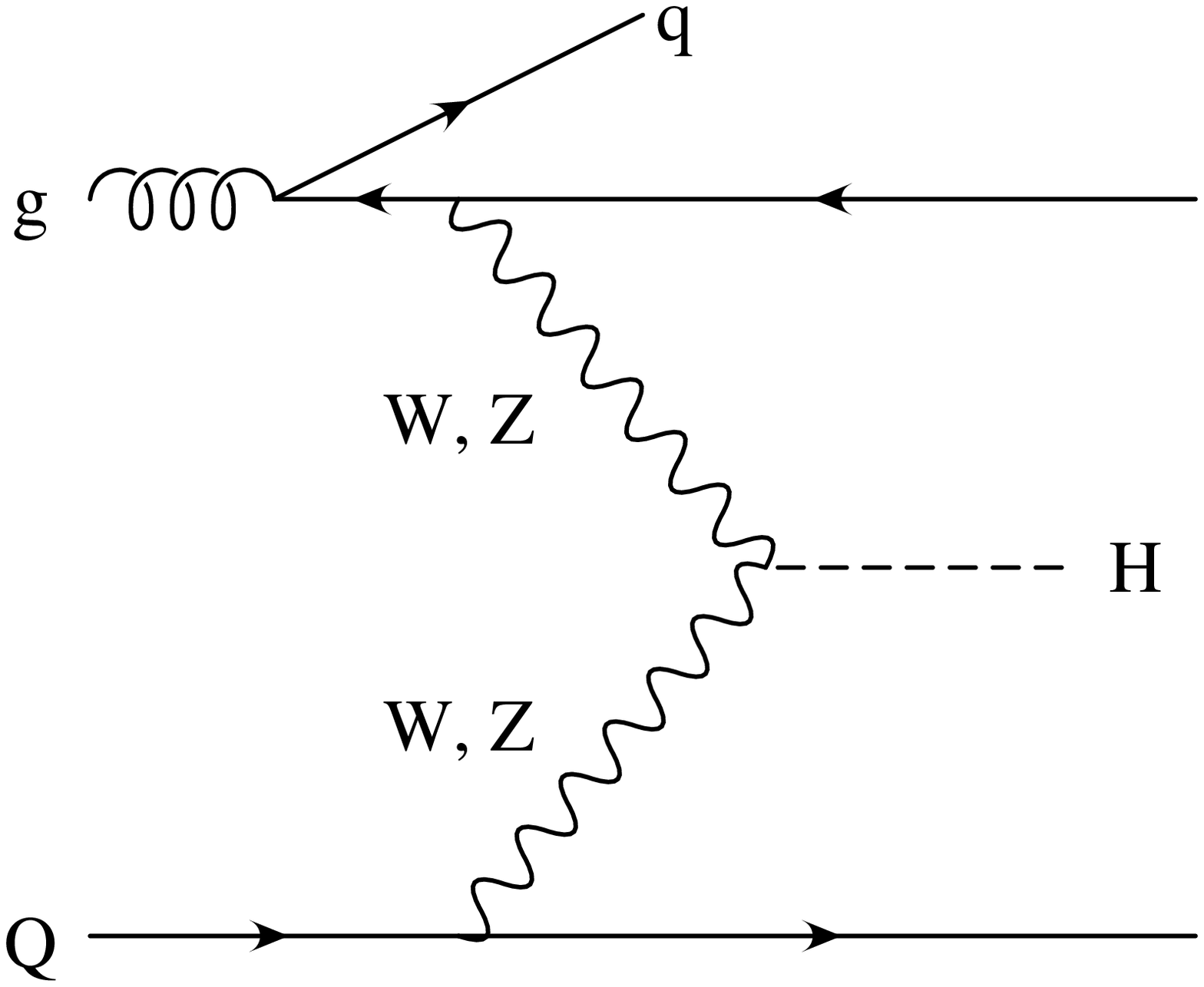,width=0.35\textwidth,clip=}}
} 
\caption{Sample of real-emission contributions to Higgs boson production via
vector-boson fusion. Corrections for the upper quark line only are shown:
final-state gluon radiation~(a) and initial-state radiation process~(b).}
\label{fig:real}
\end{figure}
At NLO, all the vertex corrections (see an example in
fig.~\ref{fig:born_virt}(b)) and all the real emission diagrams (see
fig.~\ref{fig:real} for a couple of representative diagrams) must be
included. Because of the color singlet nature of the exchanged weak boson,
any interference term between sub-amplitudes with gluons attached to both the
upper and the lower quark lines vanishes identically at order $\alpha_s$.

We have computed all the amplitudes numerically, using the helicity-amplitude
formalism of refs.~\cite{Hagiwara:1985yu,Hagiwara:1988pp}, in a similar way to
what was done in ref.~\cite{Figy:2003nv}.

The virtual corrections are particularly simple, since they factorize on the
Born squared amplitude ${\cal B}$. Following the notation of eq.~(2.92) of
ref.~\cite{Frixione:2007vw}, the finite part of the virtual corrections, once
an appropriate normalization term is factorized in front, is given by
\beq
{\cal V}_{\rm fin} =
\CF \lq-\log^2\(-\frac{\mur^2}{q_1^2}\)-3\log\(-\frac{\mur^2}{q_1^2}\)
-\log^2\(-\frac{\mur^2}{q_2^2}\)-3\log\(-\frac{\mur^2}{q_2^2}\) +  
2 c_v  \rq {\cal B}\,,
\end{equation}
where $\mur$ is the renormalization scale, $q_1$ and $q_2$ are the
space-like momenta of the two exchanged weak bosons, and $c_v=-8$
in dimensional regularization. Since there are no external gluons at the Born
level, all the spin correlated Born amplitudes ${\cal B}_{\mu\nu}$ are
zero.  From color conservation along the upper or lower line, we can
easily derive that the symmetric matrix  of color correlated Born cross
section ${\cal B}_{ij}$ has elements
\begin{equation}
{\cal B}_{14} ={\cal B}_{25} = \CF \,{\cal B}\,,
\end{equation}
all other elements being zero.

The Born total cross section is finite. For this reason we could generate
events with no cuts at the partonic level at leading order~(LO). The Born
phase space then covers the entire phase space. Although, in principle, VBF
generation cuts could be applied, we had no need to do that since a large
number of events can be easily generated.

Since there is only one Feynman diagram at Born level with only quarks, the
assignment of color flow is straightforward and unambiguous, and follows
directly the propagation of quarks and/or antiquarks.

\subsection{Tagging parton lines}
\begin{figure}[tbh]
\begin{center}
  \epsfig{file=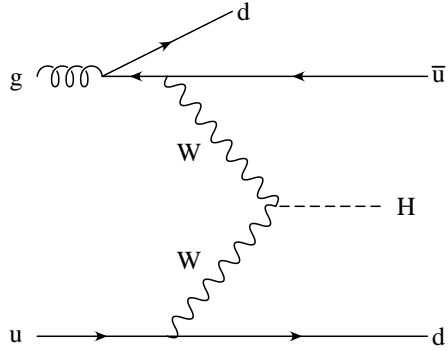,width=0.4\textwidth}
\end{center}
\caption{\label{fig:wwfus} Example of NLO gluon-initiated correction to Higgs
  boson production in VBF: $ g u\rightarrow H \bar{u} d d$.}
\end{figure}

\begin{figure}[tbh]
\begin{center}
  \epsfig{file=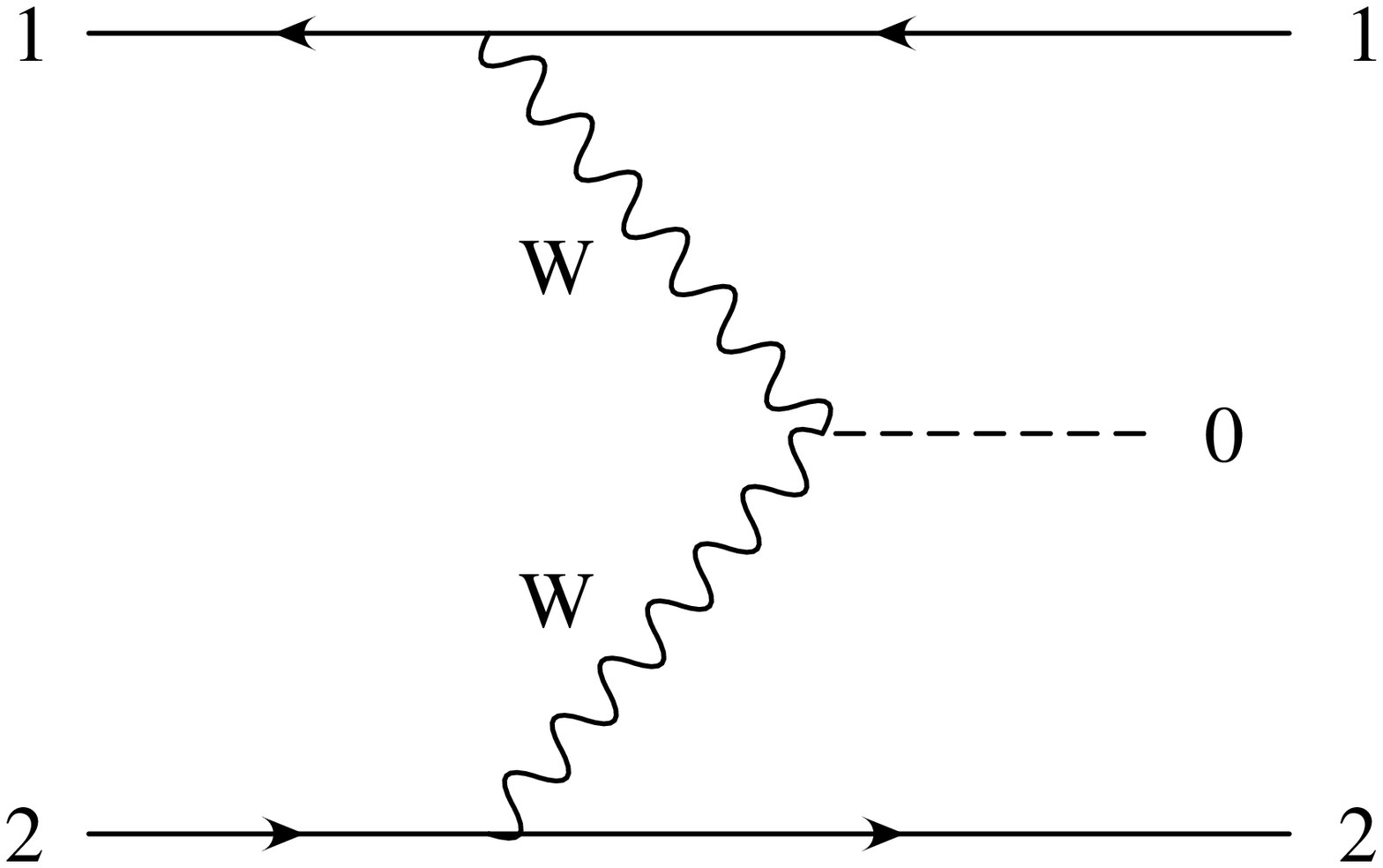,width=0.35\textwidth}\qquad\quad
  \epsfig{file=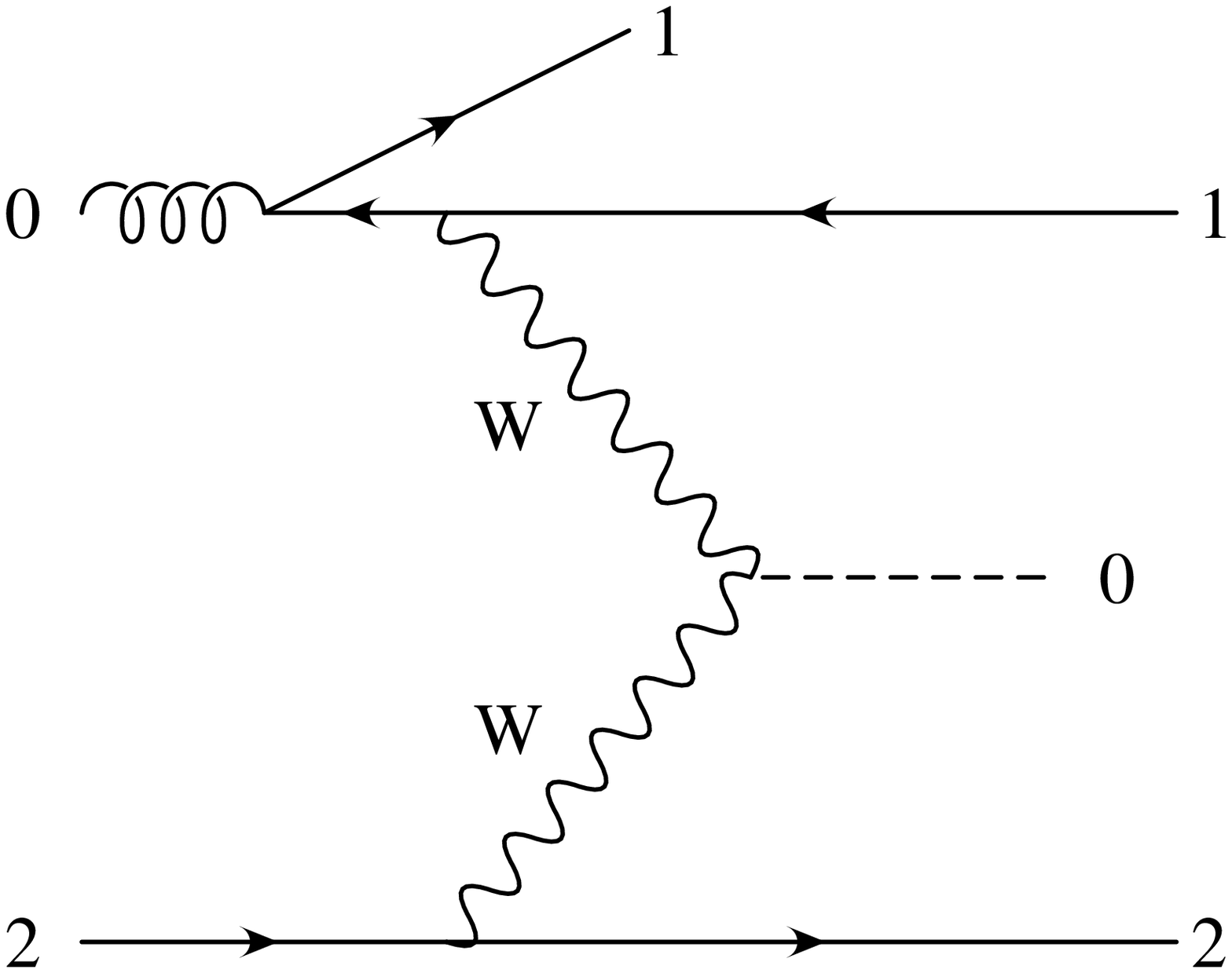,width=0.35\textwidth}
\end{center}
  \caption{\label{fig:wwfustags} Tag assignment for the underlying Born graph
  $\bar d u \to H \bar u d$ and its gluon-initiated real diagram $ g u\to H
  \bar{u} d d$. }
\end{figure}

In Higgs boson production via VBF, it is convenient to treat the upper quark
line as distinct from the lower one (see fig.~\ref{fig:born_virt}).  In fact,
radiation from the upper quark line has no interference with radiation from
the lower line, due to color flow (colorless particle exchanged in the $t$
channel). Since the \POWHEGBOX{} searches for radiation regions automatically,
it does not in principle consider the upper and lower VBF lines as distinct.
Consider for example the real graph depicted in fig.~\ref{fig:wwfus}.
It corresponds to a gluon-initiated next-to-leading correction to VBF Higgs
boson production: $g u \rightarrow H \bar{u} d d$. It is clear that
the two $d$ quarks in the final state have a very different role, and should
be kept distinct. However, as far as the flavour combinatorics is concerned,
they are considered identical in the \POWHEGBOX{}, that assumes that the
graphs are already symmetrized with respect to identical final-state
particles. Thus, the combinatoric algorithm will generate two regions for
this graph, corresponding to either $d$ being collinear to the incoming
gluon. In order to overcome this problem, the \POWHEGBOX{} allows the
possibility to attribute a tag to each line, so that lines with the same
flavour but different tags will be treated differently from the combinatoric
point of view. In the example at hand, one assigns the tags according to the
scheme in figure~\ref{fig:wwfustags}. We arbitrarily assign a tag equal to
zero to particles that we do not need to tag (the initial-state gluon and the
produced Higgs boson).
Within this scheme of assigning the tags, the two final-state $d$ quarks will
be treated as different from the combinatoric point of view. Only the quark
tagged as $1$ in the real graph will generate a singular region, since if
quark 2 were collinear to the incoming gluon, the associated underlying Born
would have an incoming antiquark tagged as 2, and thus would not be present.

Once the different singular regions have been found, the tag has no further
use and the \POWHEGBOX{} proceeds in the generation of radiation in the usual
way (see ref.~\cite{Frixione:2007vw,POWHEGBOX}).

\subsection{Tuning the real cross section in \POWHEG{}}
In \POWHEG{} it is possible to tune the contribution to the real cross
section that is treated with the Monte Carlo shower technique.  This was
pointed out first in ref.~\cite{Nason:2004rx}, where the \POWHEG{} method was
formulated, and it was first implemented in ref.~\cite{Alioli:2008tz}.
In \POWHEG{} there is the possibility to separate the real cross section, in
a given singular region $\alpha$, as follows
\begin{equation}
\label{eq:singplusreg1}
  R^{\alpha} = R^{\alpha}_s + R^{\alpha}_f, 
\end{equation}
where $R^{\alpha}_f$ has no singularities and only $R^{\alpha}_s$ is singular
in the corresponding region. In practice, the separation may be achieved, for
example, using a function of the transverse momentum of the radiation $0
\leqslant F\! \left( k_T^2 \right) \leqslant 1$, that approaches 1 when its
argument vanishes, and define
\begin{eqnarray}
  R^{\alpha}_s & = & R^{\alpha} F\! \left( k_T^2 \right), \\
  R^{\alpha}_f & = & R^{\alpha}  \lq 1 - F\! \left( k_T^2 \right) \rq . 
\end{eqnarray}
One carries out the whole \POWHEG-style generation using
$R^{\alpha}_s$ rather than $R^{\alpha}$. The contribution $R^{\alpha_{}}_f$,
being finite, is generated with standard NLO techniques, and fed into a shower
Monte Carlo as is. This feature is implemented in the \POWHEGBOX.

More generally $F$ can be chosen as a general function of the kinematic
variables, provided it approaches 1 in the singular region. This turns out to
be useful in all cases when the ratio $R/B$ (real over Born cross section) in
the \POWHEG{} Sudakov exponent becomes too much larger than its corresponding
collinear or soft approximation (see for example
ref.~\cite{Alioli:2008gx}). In this case, radiation generation becomes highly
inefficient. In Higgs boson production via VBF, we have chosen the function
$F$ in the following way: if the real squared amplitude (no parton
distribution functions included), in a particular singular region, is greater
than five times its soft and collinear approximation, then $F$ is set to
zero, otherwise is set to one. This theta-function type choice could be also
made milder, but we have found it to work well in this case.  We also stress
that this procedure remedies automatically to the Born zeros problem examined
in ref.~\cite{Alioli:2008gx}.

\subsection{The CKM-matrix treatment}
\label{sec:CKM}
In the calculation of the partonic matrix elements, all partons have been
treated as massless.  This gives rise to a different treatment of quark
flavours for diagrams where a $Z$ boson or a $W$ boson is exchanged in the $t$
channel.  In fact, for all $Z$-exchange contributions, the $b$-quark is
included as an initial and/or final-state massless parton. The $b$-quark
contributions are quite small, however, affecting the Higgs boson production
cross section at the $3$\% level only.  For $W$-exchange contributions, no
initial $b$-quark has been considered, since it would have produced mostly a
$t$ quark in the final state, that would have been misleadingly treated as
massless.

In the \POWHEG{} generation of events, we have used a diagonal form (equal to
the identity matrix) for the Cabibbo-Kobayashi-Maskawa~(CKM) matrix $V_{\rm
\sss CKM}$. In this way we reduce the number of possible Feynman diagrams
(and consequently of singular regions) to a minimal set.  At this stage, this
approximation is not a limitation of our calculation, as long as no
final-state quark flavour is tagged (no $c$ tagging is done, for example). In
fact the sum over all flavours, using the exact $V_{\rm \sss CKM}$, is
equivalent to the result obtained using the identity matrix, due to 
unitarity.

At the end of the generation of radiation done by \POWHEG, just
before the event is showered by a specific shower Monte Carlo program, the
final state quarks are ``reweighted'' according to the following CKM matrix
\begin{equation}
\begin{array}{c}
\\
V_{\rm \sss CKM}=
\end{array}
\begin{array}{c c}
& d\quad\quad\ \ s\ \ \quad\quad b \\
\begin{array}{c}
u\\
c\\
t
\end{array} 
&
\left(
\begin{array}{c c c}
0.9748 & 0.2225 & 0.0036\\
0.2225 & 0.9740 & 0.041\\
0.009 & 0.0405 & 0.9992
\end{array}
\right),
\end{array}
\end{equation}
and the final-state quark flavour is changed accordingly.  
An example will illustrate this issue more clearly. Let's suppose that we
have generated a \POWHEG{} kinematics for the subprocess $u s \to H d c g$,
using an identity matrix for $V_{\rm \sss CKM}$.  We first concentrate on the
upper leg, where we have the decay $u \to W^+ d$. We retain the $d$-quark
flavour in the final state with probability equal to $V_{ud}^2/\Sigma$, where
$\Sigma =V_{ud}^2+V_{us}^2+V_{ub}^2$, while we change it to an $s$-quark or
$b$-quark flavour according to the probabilities $V_{us}^2/\Sigma$ and
$V_{ub}^2/\Sigma$, respectively.  The lower line undergoes a similar
treatment, but for the $s \to W^- c$ subprocess: the final-state $c$ quark is
retained with a probability equal to $V_{cs}^2/\Sigma'$, where
$\Sigma'=V_{us}^2+V_{cs}^2+V_{ts}^2$, while it is changed into a $u$ or a $t$
quark with probability equal to $V_{us}^2/\Sigma'$ and $V_{ts}^2/\Sigma'$,
respectively. The case of a $t$-quark production is, in all cases, disregarded
at the end.

\section{Results}
\label{sec:results}
In this section we present comparisons of the fixed order next-to-leading
calculation and the results obtained after the shower performed by
\HERWIG{}~6.510 and \PYTHIA~6.4.21.
We have used the CTEQ6M~\cite{Pumplin:2002vw} set
for the parton distribution functions and the associated value of
\mbox{$\Lambda_{\scriptscriptstyle\overline{\rm
MS}}^{(5)}=0.226$~GeV}.  Furthermore, as discussed in
refs.~\cite{Frixione:2007vw,Nason:2006hfa}, we use a rescaled value
\mbox{$\Lambda_{\scriptscriptstyle\rm{MC}}=
1.569\,\Lambda_{\scriptscriptstyle\overline{\rm MS}}^{(5)}$} in the
expression for $\as$ appearing in the Sudakov form factors, in order
to achieve next-to-leading logarithmic accuracy.

Although the matrix-element calculation has been performed in the
massless-quark limit, the lower cutoff
in the generation of the radiation has been fixed according to the mass of
the emitting quark. The lower bound on the transverse momentum for the
emission off a massless emitter ($u$, $d$, $s$) has been set to the value
$\ptmin = \sqrt{5}\,\Lambda_{\scriptscriptstyle\rm{MC}}$. We instead choose
$\ptmin$ equal to $m_c$ or $m_b$ when the gluon is emitted by a charm or a
bottom quark, respectively.  We set $m_c=1.55$~GeV and $m_b= 4.95$~GeV.

The renormalization $\mur$ and factorization $\muf$ scales have been taken
equal to the transverse momentum of the radiated parton
during the generation of radiation,
as the \POWHEG{} method requires. The transverse momentum of the radiated
parton is taken, in the case of initial-state radiation, as exactly
equal to the transverse momentum of the parton with respect to the
beam axis. For final-state radiation one takes instead
\begin{equation}
\pt^2=2E^2(1-\cos\theta),
\end{equation}
where $E$ is the energy of the radiated parton and $\theta$
the angle it forms with respect to the final-state parton that has
emitted it, both taken in the partonic center-of-mass frame.

We have also taken into account properly the heavy-flavour thresholds in the
running of $\as$ and in the parton distribution functions~(pdf's), by
changing the number of active flavours when the renormalization or
factorization scales cross a mass threshold.  In the $\bar{B}$ calculation,
instead, $\mur$ and $\muf$ have been chosen equal to the Higgs boson mass,
whose value has been fixed to $m_H=120$~GeV, and its corresponding width
$\Gamma_H = 0.00437$~GeV.  The other relevant parameters are
\begin{equation}
M_W= 79.964 \mbox{ GeV}\,,\quad M_Z= 91.188 \mbox{ GeV}\,,
\quad \sin^2\theta_W^{\rm eff}=0.23102\,,
\quad \alpha^{-1}_{\rm em}(M_Z)=128.930\,,
\end{equation}
and we have also set $\Gamma_W=\Gamma_Z=0$ in all the propagators.  
From the above values, the weak coupling has been computed as
$g=\sqrt{4\pi\alpha_{\rm em}}/\sin\theta_W^{\rm eff}$.

Using the \POWHEGBOX, we have generated 500000 events that we have interfaced
both with \HERWIG{} and with \PYTHIA, for an energy of 14~TeV at the LHC $pp$
collider. Events with a top quark in the final state have been neglected, for
the reasons discussed in section~\ref{sec:CKM}.

The defining feature of weak-boson fusion events at hadron colliders is the
presence of two forward tagging jets, which, at LO, correspond to the two
scattered quarks in the process $\bar qQ\to H \bar q Q$. Their observation,
in addition to exploiting the properties of the Higgs boson decay products,
is crucial for the suppression of
backgrounds~\cite{Rainwater:1998kj,Plehn:1999xi,Rainwater:1999sd,
  Kauer:2000hi,Rainwater:1997dg,DelDuca:2001eu,DelDuca:2001fn}. The stringent
acceptance requirements on the tagging jets imply that their distributions
must be known precisely for a reliable prediction of the SM Higgs signal
rate. We remind the reader that comparison of the observed Higgs boson
production rate with this SM cross section allows to determine its
couplings~\cite{Zeppenfeld:2000td,Duhrssen:2004cv}.  The theoretical error on
the cross section thus directly feeds into the uncertainty of measured
couplings.

Since the Higgs boson is a scalar, it does not induce any spin correlation in
its decay products. We concentrate then only on the analysis of tagging-jet
distributions and we do not impose cuts on the Higgs boson decay products.

After the shower, the final state consists of a Higgs boson plus a number of
jets originating from the \POWHEG{} hard partons and from the shower.  Jets
are defined according to the $\kt$ algorithm~\cite{Catani:1993hr}, as
implemented in the {\tt FASTJET} package~\cite{Cacciari:2005hq}, setting
$R=0.7$ and imposing the following cuts on their transverse momentum and
rapidity
\begin{equation}
\label{eq:minimal_cuts}
\pt{}_j > 20~{\rm GeV}, \qquad  |y_j| < 5\,.
\end{equation}
The two tagging jets are the two jets with highest $\pt$.  They must
satisfy the additional constraints 
\beq
\label{eq:VBF_cuts}
\pt^{\rm tag} > 30~{\rm GeV}, \qquad 
|y_{j_1} - y_{j_2}| > 4.2\,, \qquad y_{j_1}\cdot y_{j_2} < 0\,, \qquad
m_{jj} > 600~{\rm GeV}\,,
\end{equation}
i.e.~they must be well separated in rapidity, lie in opposite hemispheres
and have a large invariant mass.

Only one quarter of the whole number of generated events passes all the cuts,
and we have used only these events for the following analysis.

\begin{figure}[htb]
\begin{center}
\epsfig{file=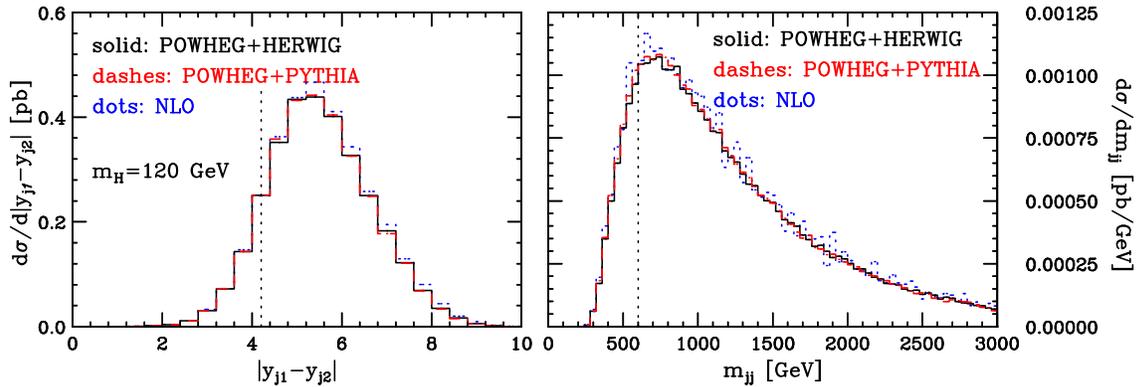,width=\figwidth}
\end{center}

\captskip
\caption{\label{fig:dely_mjj} Differential cross section as function of the
  absolute value of the difference of the rapidity of the two tagging jets
  (left panel) and of their invariant mass (right panel). In the left panel,
  we have excluded the $|y_{j_1} - y_{j_2}| > 4.2$ cut of
  eq.~(\ref{eq:VBF_cuts}), while in the right panel we have excluded the
  $m_{jj} > 600$~GeV constraint. The black dotted line marks the position of
  these cuts.}
\end{figure}
In fig.~\ref{fig:dely_mjj} we present the differential cross section as
function of the absolute value of the difference of the rapidity of the two
tagging jets and of their invariant mass.  The characteristic features of the
VBF Higgs boson production are left unchanged by the shower: the differential
cross section has a peak around 5.5 in the absolute difference of the
rapidity of the two tagging jets and the invariant dijet mass still peaks for
values around 700~GeV. The NLO curve and the \POWHEG{} results, after the
\HERWIG{} and \PYTHIA{} shower, are almost indistinguishable.  This result
validates the use of the two main cuts (on the difference in rapidity and on
the invariant dijet mass) as crucial selection criteria to reduce the
background to this process.

\begin{figure}[htb]
\begin{center}
\epsfig{file=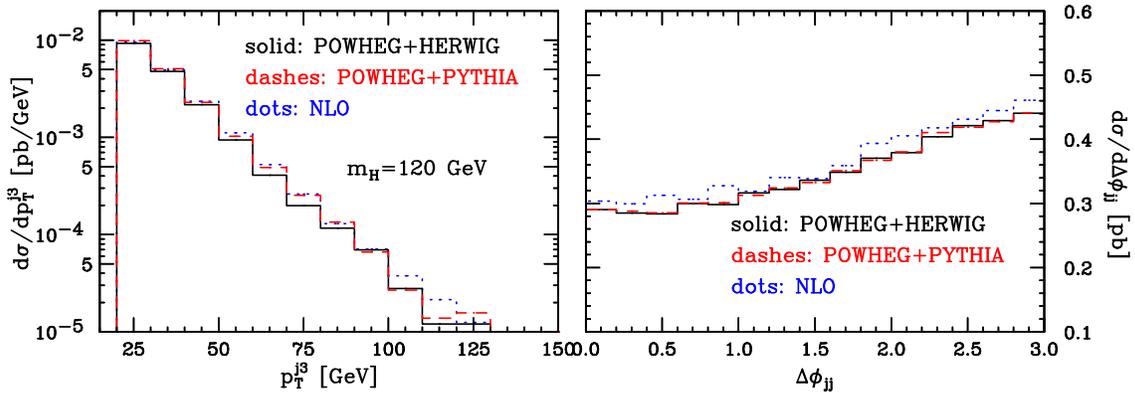,width=\figwidth}
\end{center}
\captskip
\caption{\label{fig:pt3_phijj} Transverse momentum ($\pt^{j_3}$) distribution
of the third hardest jet (left panel) and azimuthal-distance distribution of
the two tagging jets, $\Delta\phi_{jj}$ (right panel).}
\end{figure}
In fig.~\ref{fig:pt3_phijj} we plot the transverse momentum of the third
hardest jet (left panel) and the azimuthal distance between the two tagging
jets, $\Delta\phi_{jj}$ (right panel). This last quantity is of particular
interest since it is sensitive to the CP nature of the Higgs boson
couplings~\cite{Plehn:2001nj,Hankele:2006ja,Klamke:2007cu,Hankele:2006ma}. For
example, in gluon-fusion Higgs boson production plus two jets, the analysis
of the azimuthal-angle correlations provides for a direct measurement of the
CP properties of the $Ht\bar{t}$ Yukawa coupling which is responsible for the
effective $Hgg$ vertex.  In the VBF process $q(p_1) \, Q(p_2) \to
H(p_3)\, q(p_4)\,Q(p_5)$, the matrix element squared is proportional to
\begin{equation} 
|{\cal A}_{\rm VBF}|^2 \propto \frac{1}{\(2 \, p_1\cdot p_4+M_V^2\)^2}\,
 \frac{1}{\(2 \, p_2\cdot p_5+M_V^2\)^2} \, \hat{s} \, m_{jj}^2\;,
\end{equation}
where $M_V$ is the mass of the exchanged $t$-channel vector boson, and is
dominated by the contribution in the forward region, where the dot-products
in the denominator are small.  Since the dependence of $m_{jj}^2$ on
$\Delta \phi_{jj}$ is mild, we have the flat behavior depicted in
fig.~\ref{fig:pt3_phijj}.
Good agreement is found in the two \POWHEG{} results and both agree with the
NLO differential cross section.

\begin{figure}[htb]
\begin{center}
\epsfig{file=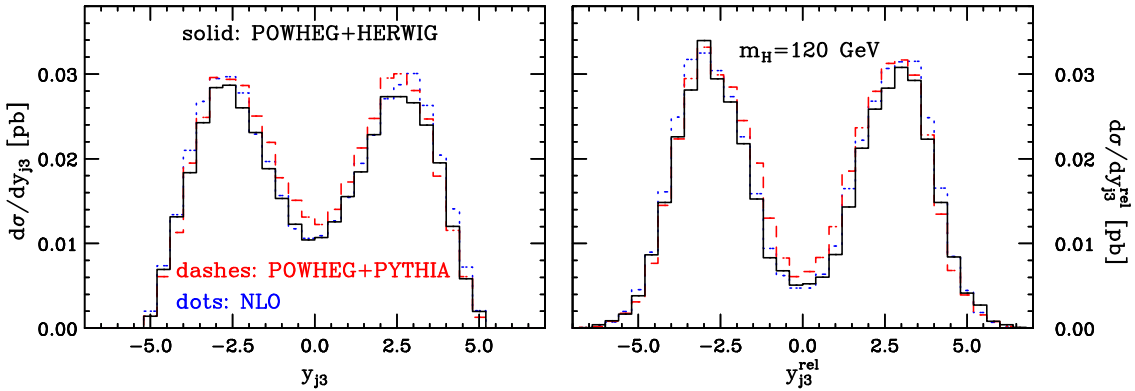,width=\figwidth}
\end{center}
\captskip
\caption{\label{fig:y3_rel} Rapidity $y_{j_3}$ of the third hardest jet (the
one with highest $\pt$ after the two tagging jets) on the left panel and
rapidity of the same jet with respect to the average of the rapidities of the
two tagging jets $y_{j_3}^{\rm rel} = y_{j_3} - \(y_{j_1}+y_{j_2}\)/2$ on the
right panel.}
\end{figure}
An additional feature characterizing VBF Higgs boson production is the fact
that, at leading order, no colored particle is exchanged in the $t$ channel
so that no $t$-channel gluon exchange is possible at NLO, once we neglect, as
stated in section~\ref{sec:powhegbox}, the small contribution due to
equal-flavour quark scattering with $t\leftrightarrow u$ interference.  The
different gluon radiation pattern expected for Higgs boson production via VBF
compared to its major backgrounds ($t\bar{t}$ production, QCD $WW + 2$~jet
and QCD $Z+2$~jet production) is at the core of the central-jet veto proposal,
both for light~\cite{Kauer:2000hi} and heavy~\cite{Barger:1994zq} Higgs boson
searches.  A veto of any additional jet activity in the central-rapidity
region is expected to suppress the backgrounds more than the signal, because
the QCD backgrounds are characterized by quark or gluon exchange in the
$t$-channel. The exchanged partons, being colored, are expected to radiate
off more gluons.

For the analysis of the Higgs boson coupling to gauge bosons, Higgs
boson~+~2~jet production via gluon fusion may also be treated as a background
to VBF.  When the two jets are separated by a large rapidity interval, the
scattering process is dominated by gluon exchange in the $t$-channel.
Therefore, like for the QCD backgrounds, the bremsstrahlung radiation is
expected to occur everywhere in rapidity. An analogous difference in the
gluon radiation pattern is expected in $Z+2$~jet production via VBF fusion
versus QCD production~\cite{Rainwater:1996ud}.  In order to analyze this
feature, in ref.~\cite{DelDuca:2004wt} the distribution in rapidity of the
third jet was considered in Higgs +~3~jet production via VBF and via gluon
fusion, using cuts similar to the ones in eqs.~(\ref{eq:minimal_cuts})
and~(\ref{eq:VBF_cuts}).  The analysis was done at the parton level only. It
showed that, in VBF, the third jet prefers to be emitted close to one of the
tagging jets, while, in gluon fusion, it is emitted anywhere in the rapidity
region between the tagging jets. Thus, at least with regard to the hard
radiation of a third jet, the analysis of
refs.~\cite{DelDuca:2004wt,DelDuca:2006hk,Andersen:2008gc} confirmed the
general expectations about the bremsstrahlung patterns in Higgs production
via VBF versus gluon fusion.

To study the distribution of the third hardest jet (the one with highest
$\pt$ after the two tagging jets), we plot in fig.~\ref{fig:y3_rel} its
rapidity and its rapidity with respect to the average of the rapidities of
the two tagging jets 
\beq 
y_{j_3}^{\rm rel} = y_{j_3} - \frac{y_{j_1}+y_{j_2}}{2}\,.  
\eeq
The distributions obtained using \POWHEG{} interfaced to \HERWIG{} and
\PYTHIA{}, are very similar and turn out to be well modeled by the respective
distributions of the NLO jet: the third jet generally tends to be emitted in
the vicinity of either of the tagging jets.

\begin{figure}[htb]
\begin{center}
\epsfig{file=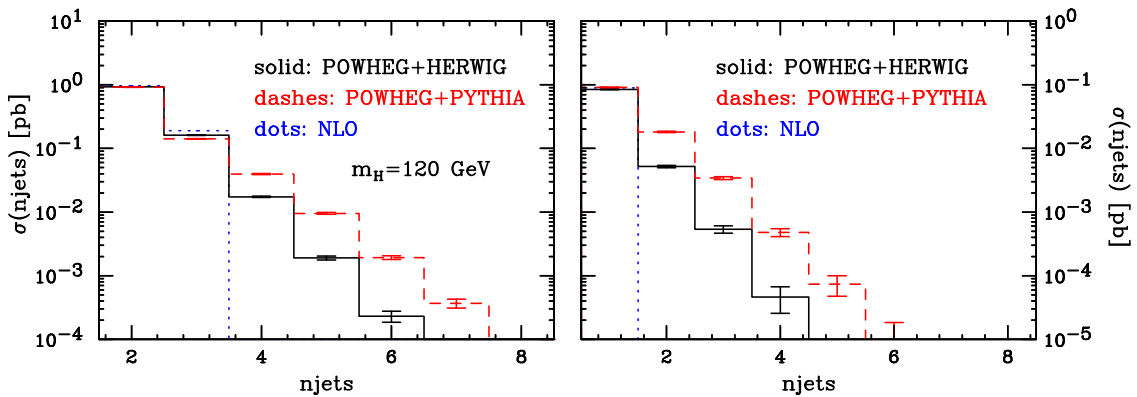,width=\figwidth}
\end{center}
\captskip
\caption{\label{fig:jet_mult} Jet-multiplicity distribution for jets that
  pass the cuts of eqs.~(\ref{eq:minimal_cuts}) and~(\ref{eq:VBF_cuts})
  (left panel) and those that fall within
  the rapidity interval of the two tagging jets, $\min\(y_{j_1},y_{j_2}\) <
  y_j < \max\(y_{j_1},y_{j_2}\)$ (right panel). }
\end{figure}
In order to quantify the jet activity, we plot the jet-multiplicity
distribution for jets that pass the cuts of eqs.~(\ref{eq:minimal_cuts})
and~(\ref{eq:VBF_cuts}) in the left panel of fig.~\ref{fig:jet_mult}. Again,
the first two tagging jets and the third jet are well represented by the NLO
cross section, that obviously cannot contribute to events with more than three
jets.  From the 4th jet on, the showers of \HERWIG{} and \PYTHIA{} produce
sizable differences (note the log scale of the plot), the jets from
\PYTHIA{} being harder than those from the \HERWIG{} shower.

A similar behavior is present when we investigate the jet activity restricted
in the rapidity interval between the tagging jets.  In the right panel of
fig.~\ref{fig:jet_mult} we plot the jet-multiplicity distribution for jets
that fall within the rapidity interval of the two tagging jets (also called
veto jets), i.e.
\beq
\min\(y_{j_1},y_{j_2}\) < y_j < \max\(y_{j_1},y_{j_2}\).
\eeq
No difference is seen for the NLO jet with respect to the results of
\POWHEG{} interfaced to \HERWIG{} and \PYTHIA{}, while for
multiplicity greater than one, \PYTHIA{} generates harder jets if compared to
\HERWIG.

\begin{figure}[htb]
\begin{center}
\epsfig{file=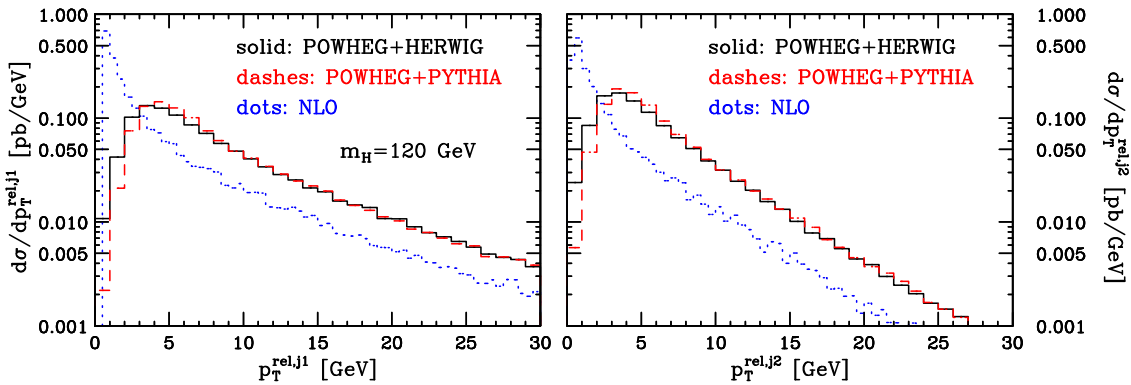,width=\figwidth}
\end{center}
\captskip
\caption{\label{fig:pt_rel}  Relative transverse momentum of all the
particles clustered inside one of the two tagging jets, in the reference
frame where that jet has zero rapidity, defined according to
eq.~(\ref{eq:def_ptrel}). In the left panel  $\pt^{{\rm rel},j_1}$ is plotted
while in the right panel we plotted $\pt^{{\rm rel},j_2}$.}
\end{figure}

Striking differences between the NLO results and \POWHEG{} can be seen, as
expected, if we consider distributions sensitive to the collinear/soft
regions, such as the relative transverse momentum of all the particles
clustered inside one of the two tagging jets: $\pt^{{\rm rel},j_1}$ and
$\pt^{{\rm rel},j_2}$.  This quantity is defined as follows:
\begin{itemize}
\item[-] for each of the two tagging jets, we perform a longitudinal boost to
  a frame where the jet has zero rapidity.
\item[-] In this frame, we compute the quantity
\begin{equation}
\label{eq:def_ptrel}
\pt^{{\rm rel},j}=\sum_{i\in j} \frac{| \vec{k}^i \times
  \vec{p}^{\, j}|}{|\vec{p}^{\, j}|}\,,
\end{equation}
where $k^i$'s are the momenta of the particles that belong to the jet
that, in this frame, has momentum $p^{\,j}$.
\end{itemize}
This quantity is thus the sum of the absolute values of the transverse
momenta, taken with respect to the jet axis, of the particles inside the
hardest jet, in the frame specified above.

As last comparison, we have studied the probability of finding a veto jet and
we have compared this with the results of ref.~\cite{Figy:2007kv}.  
For the central jet veto proposal, events are discarded if any additional jet
with a transverse momentum above a minimal value, $\pt{}_{\rm ,veto}$, is
found between the tagging jets.  In ref.~\cite{Figy:2007kv}, the authors
present a calculation of the dominant NLO correction to the production of a
Higgs boson plus three jets, i.e.~the LO of this correction coincides with
the real contribution in this paper, since here we deal with NLO
correction to Higgs boson production plus two jets.

In order to make a comparison with the results of ref.~\cite{Figy:2007kv}, we
need to slightly adjust our cuts to the ones used in that paper. The cuts in
eqs.~(\ref{eq:minimal_cuts}) and~(\ref{eq:VBF_cuts}) are replaced by the
following 
\begin{equation}
\pt{}_j > 20~{\rm GeV}, \qquad  |y_j| < 4.5\,,
\end{equation}
and
\begin{equation}
\pt^{\rm tag} > 30~{\rm GeV}, \qquad 
|y_{j_1} - y_{j_2}| > 4\,, \qquad y_{j_1}\cdot y_{j_2} < 0\,, \qquad
m_{jj} > 600~{\rm GeV}\,,
\end{equation}
with jets reconstructed with resolution parameter $R=0.8$.  The Higgs boson
decay products (generically called ``leptons" in the following) are required
to fall between the two tagging jets in rapidity and they should be well
observable. While the exact criteria for the Higgs decay products will depend
on the channel considered, such specific requirements here are substituted by
generating isotropic Higgs boson decay into two massless ``leptons'' (which
represent $\tau^{+} \tau^{-}$ or $\gamma \gamma$ final states) and requiring
\begin{equation}
p_{T\ell} \geq 20~{\rm GeV} \,,\qquad |\eta_{\ell}| \leq 2.5 \,,\qquad
\triangle R_{j\ell} \geq 0.6 \, , 
\end{equation}
where $\triangle R_{j\ell}$ denotes the jet-lepton separation in the
rapidity-azimuthal angle plane. In addition, the two ``leptons'' are required
to fall between the two tagging jets in rapidity
\begin{equation}
\min\(y_{j_1},y_{j_2}\)+0.6<\eta_{\ell_{1,2}}<\max\(y_{j_1},y_{j_2}\)-0.6\,.
\end{equation}
Note that no reduction due to branching ratios for specific final states
has been included in the calculation. 
In addition we set $\muf=\mur=40$~GeV, since this value minimizes
the scale dependence of the NLO $Hjjj$ prediction and, at the same time, it
provides optimal agreement between LO and NLO $Hjjj$ total cross sections,
within the VBF cuts.

\begin{figure}[htb]
\begin{center}
\epsfig{file=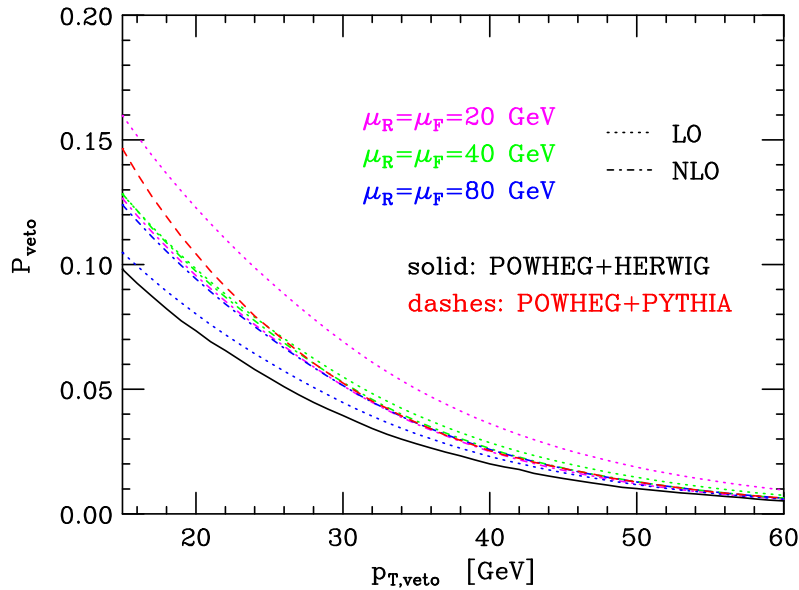,width=0.7\figwidth}
\end{center}
\captskip
\caption{\label{fig:ptveto} Probability of finding a veto jet defined as in
  eq.~(\ref{eq:P_veto}). The dotted curves depict the LO results and the
  dotdashed ones the NLO results for $Hjjj$ production, taken from
  ref.~\cite{Figy:2007kv}, for $\muf=\mur=20$~GeV (magenta),
  $\muf=\mur=40$~GeV (green) and $\muf=\mur=80$~GeV (blue). The solid black
  and dashed red curves represent the results of \POWHEG{} interfaced to
  \HERWIG{} and \PYTHIA{}.}
\end{figure}

We are now in the position to make a comparison between our results and the
ones in ref.~\cite{Figy:2007kv} for the the probability, $P_{\rm veto}$, of
finding a veto jet
\begin{equation}
\label{eq:P_veto}
P_{\rm veto} = \frac{1}{\sigma_2^{\sss NLO}} \int_{\pt{}_{\rm ,veto}}^{\infty} 
d\pt^{\,j,{\rm veto}} \frac{d\sigma}{d\pt^{\,j,{\rm veto}}}\,,
\end{equation}
where $\pt^{\,j,{\rm veto}}$ is the transverse momentum of the hardest veto
jet, and $\sigma_2^{\sss NLO}=0.723$~pb is the total cross section (within
VBF cuts) for $Hjj$ production at NLO, computed with $\muf=\mur=m_H$.  If
fig.~\ref{fig:ptveto} we have plotted the LO and NLO $Hjjj$ curves of
ref.~\cite{Figy:2007kv}, and the results we have obtained with \POWHEG.  We
see that the results of \POWHEG{} interfaced to \HERWIG{} and \PYTHIA{} are
consistent with the LO band obtained with a change of the renormalization and
factorization scale by a factor of two.  In addition, notice that the
distance between \POWHEG{}+\PYTHIA{} and \POWHEG{}+\HERWIG{} is comparable to
the scale uncertainty of the leading order result.

\section{Conclusions}
\label{sec:conc}
In this paper we have described a complete implementation of Higgs boson
production in vector-boson fusion at next-to-leading order in QCD, in the
\POWHEG{} framework.  Together with $Z+1$~jet production~\cite{POWHEG_Zjet},
this is the first time that the NLO results for these processes are merged
with a shower.  The actual implementation is based on the \POWHEGBOX{}
package: this is an automated package designed to allow the construction
of a \POWHEG{} implementation for any given NLO calculation.
New features with respect to previous implementations of \POWHEG{} have
been applied to deal with VBF Higgs boson production: the tagging of
parton lines and the tuning of the real cross section in \POWHEG{}.

%
We have shown and discussed several distributions after imposing typical VBF
cuts (the two jets with highest transverse momentum must be well separated in
rapidity, lie in opposite hemispheres and have a large invariant mass).  These
cuts strongly suppress
many backgrounds to Higgs boson production via VBF, at the
LHC.

We have showered the \POWHEG{} outputs both with \HERWIG{} and with \PYTHIA{}
and found good agreement between the two showers.  Furthermore, we have found
that overall our calculation confirms results obtained previously with
parton-level Monte Carlo programs.  We found some discrepancies between our
\POWHEG{} result showered with \HERWIG{} and with \PYTHIA{} in the
multiplicity of final-state jets, for more than three jets.

The computer code for the \POWHEG{} implementation presented in this paper
will soon be available at the site
\url{http://moby.mib.infn.it/~nason/POWHEG}, as part of the \POWHEGBOX{}
package.

\bibliography{paper}

\end{document}